\documentclass[twocolumn]{article}

\begin{document}

\title{\bfseries
        \vspace*{-0.3in}%
\begin{flushright}\large AIAA-99-2144
\end{flushright}        \vspace*{0.1in}{\large \textbf{PROPULSION THROUGH
ELECTROMAGNETIC SELF-SUSTAINED ACCELERATION}} }
\author{Vesselin Petkov \\
Physics Department, Concordia University\\
1455 De Maisonneuve Boulevard West\\
Montreal, Quebec, Canada H3G 1M8\\
E-mail: vpetkov@alcor.concordia.ca}
\date{}
\maketitle

\begin{abstract}
As is known the repulsion of the volume elements of an uniformly
accelerating charge or a charge supported in an uniform gravitational field
accounts for the electromagnetic contribution to the charge's inertial and
gravitational mass, respectively. This means that the mutual repulsion of
the volume elements of the charge produces the resistance to its accelerated
motion. Conversely, the effect of electromagnetic attraction of opposite
charges enhances the accelerated motion of the charges provided that they
have been initially uniformly accelerated or supported in an uniform
gravitational field. The significance of this effect is that it constitutes
a possibility of altering inertia and gravitation.
\end{abstract}

\section*{Introduction}

In 1881 J. J. Thomson \cite{thomson} first realized that a charged particle
was more resistant to being accelerated than an otherwise identical neutral
particle. His observation marked the origin of the concept of
electromagnetic mass of charged particles. This concept was developed in a
full theory mostly by Heaviside \cite{heaviside}, Searle \cite{searle},
Lorentz \cite{lorentz04, lorentz}, Poincar\'{e} \cite{poincare05, poincare06}%
, Abraham \cite{abraham}, Fermi \cite{fermi, fermi22} and Rohrlich \cite
{rohrlich60, rohrlich}. It follows from this theory that it is the unbalanced%
\footnote{%
The mutual repulsion of two inertial like charges is completely balanced and
therefore there is no net force acting on the charges.} repulsion of the
volume elements of an accelerating charged particle that causes the
resistance to its acceleration known as inertia. Alternatively, the
unbalanced attraction of accelerating opposite charges results in further
enhancement of their acceleration. By the equivalence principle the opposite
effects of resistance to like charges' acceleration and increase of unlike
charges' acceleration resulting from the unbalanced repulsion of like
charges and the unbalanced attraction of unlike charges, respectively should
also occur when the charges are in a gravitational field. The equivalence
principle requires that these effects be present in a gravitational field as
well but does not provide any insight into what causes them there. The
answer to this question is that it is a spacetime anisotropy around massive
bodies that causes those effects. It manifests itself in the anisotropy of
the velocity of electromagnetic signals (for short - the velocity of light).
Since it is now believed that the anisotropy of the velocity of light around
massive bodies results from the curvature of spacetime, here we shall
discuss two \emph{independent} results which indicate that the correct
interpretation of general relativity is in terms of spacetime anisotropy,
not spacetime curvature. While the first result just shows that there is no
need for spacetime curvature since spacetime anisotropy \emph{alone}
accounts for all inertial and gravitational effects, the second one directly
demonstrates that the standard curved-spacetime interpretation of general
relativity contradicts the gravitational redshift experiments. The
implication that there is no spacetime curvature is crucial not only for
understanding and possible utilization of the effects discussed but also for
gaining deeper insight into the nature of inertia and gravitation. One
far-reaching consequence from the anisotropy of spacetime is that inertia
and gravitation can (at least in principle) be electromagnetically
manipulated. \newline
\indent \emph{It is the anisotropy of spacetime that causes the phenomena
traditionally called inertia and gravitation. }An analysis of the classical
electromagnetic mass theory in conjunction with general relativity leads to
the conclusion that (i) gravitational attraction is caused by the anisotropy
of spacetime around massive objects and (ii) inertia (and inertial mass) as
described in an accelerating reference frame originates from the spacetime
anisotropy in that frame \cite{phd, petkov}. The essence of this analysis is
as follows. Consider a classical\footnote{%
At present quantum mechanical treatment of the electromagnetic mass is not
possible since quantum mechanics does not offer a model for the quantum
object itself.} electron in the Earth's gravitational field. Due to the
anisotropy of the velocity of light in a non-inertial reference frame
(supported in the Earth's gravitational field) the electric field of an
electron on the Earth's surface is distorted which gives rise to a
self-force originating from the interaction of the electron charge with its
distorted electric field\footnote{%
That explanation of the origin of the self-force is another way of saying
that the force arises from the mutual unbalanced repulsion of the volume
elements of the electron charge.}. This self-force tries to force the
electron to move downwards\footnote{%
The self-force which starts to act on the electron whenever its electric
field distorts effectively resists this distortion. That is why the
self-force strives to make the electron move downwards with an acceleration $%
\mathbf{g} $ in order to compensate the spacetime anisotropy which
in turn will eliminate the distortion of the electron's electric
field.} and coincides with what is traditionally called the
gravitational force. The electric self-force is proportional to the gravitational acceleration $%
\mathbf{g}$ and the coefficient of proportionality is the mass ''attached''
to the electron's electric field which proves to be equal to the electron
mass. The anisotropy of the speed of light in the Earth's vicinity is
compensated if the electron is falling toward the Earth's surface with an
acceleration $\mathbf{g}$. In other words, the electron is falling in order
to keep its electric field not distorted. A Coulomb (not distorted) field
does not give rise to any self-force acting on the electron; that is why the
motion of a falling electron is non-resistant (inertial, or geodesic)%
\footnote{%
It is clear from here that a falling electron does not radiate since its
electric field is the Coulomb field and therefore does not contain the
radiation $r^{-1}$ terms \cite{phd}.}. If the electron is prevented from
falling it can no longer compensate the anisotropy of the speed of light,
its field distorts and as a result a self-force pulling the electron
downwards arises. The resistance which an electron offers to being
accelerated is similarly described in an accelerating reference frame as
caused by the anisotropy of the speed of light there. When the electron
accelerates its electric field distorts and the electron resists that
deformation. In such a way, given the fact that the speed of light is
anisotropic in non-inertial reference frames, all inertial and gravitational
effects (including the equivalence of inertial and gravitational mass) of
the electron and the other elementary charged particles are fully and
consistently accounted for if both the inertial and passive gravitational
mass of the elementary charged particles are entirely\footnote{%
On the one hand, the entirely electromagnetic mass of an elementary charged
particle - an electron for example - is supported by the fact that the
electromagnetic mass of the classical electron is equal to its observable
mass. On the other hand, the electromagnetic mass raises the question of
stability of the electron (what keeps its charge together). This question,
however, cannot be adequately addressed until a quantum-mechanical model of
the electron structure is obtained. An important feature of the
electromagnetic mass theory is that the stability problem does not interfere
with the derivation of the self-force (acting on a non-inertial electron)
containing the electromagnetic mass \cite{phd, petkov}. This hints that
perhaps there is no real problem with the stability of the electron (as a
future quantum mechanical model of the electron itself may find); if there
were one it would inevitably emerge in the calculation of the self-force.}
electromagnetic in origin. The gravitational attraction and inertia of all
matter can be accounted for as well if it is assumed that there are no
elementary neutral particles in nature. A direct consequence from here is
that only charged particles or particles that consists of charged
constituents possess inertial and passive gravitational mass\footnote{%
It is evident that in this case the electromagnetic mass theory predicts
zero neutrino mass and appears to be in conflict with the apparent mass of
the $Z^{0}$ boson which is involved in the neutral weak interactions. The
resolution of this apparent conflict could lead to either restricting the
electromagnetic mass theory (in a sense that not the entire mass is
electromagnetic) or reexamining the facts believed to prove (i) that the $%
Z^{0}$ boson is a fundamentally neutral particle (unlike the neutron), and
(ii) that it does possess inertial and gravitational mass if truly neutral.}%
. Stated another way, it is only elementary charges that comprise a body;
there is no such fundamental quantity as mass. This means that a body's
(inertial or passive gravitational) mass corresponds to the energy stored in
the electric fields of all elementary charged particles comprising the body.
However, the inertial and passive gravitational mass of a body manifest
themselves as such - as a measure of the body's resistance to being
accelerated - only if the body is subjected to an acceleration (kinematic or
gravitational). This resistance originates from the unbalanced mutual
repulsion of the volume elements of every elementary charged constituent of
the body. The active gravitational mass of a body proves to be also
electromagnetic originating from its charged constituents\footnote{%
Given that the inertial and passive gravitational masses are electromagnetic
the electromagnetic nature of the active gravitational mass follows
immediately since the three kinds of masses are equal.} since there is no
mass but only charges. As there is no need for spacetime curvature since all
gravitational effects are fully accounted for by the electromagnetic nature
of the passive gravitational mass and the anisotropic velocity of light in
the vicinity of a massive object \cite{phd, petkov}, it follows that it is
the object's charges (and their fields) that cause that anisotropy of
spacetime around the object.

One thing concerning the electromagnetic mass theory which is often
overlooked should be especially stressed: even if the mass is viewed as only
partly electromagnetic, as presently believed, it still follows that inertia
and gravitation are electromagnetic in origin but in part. It should be also
noted that once the fact of the partly electromagnetic origin of inertia and
gravitation is fully realized a thorough analysis of this open issue can be
carried out which will most probably lead to the result that electromagnetic
interaction is the only cause behind inertia and gravitation which are now
regarded as separate phenomena\footnote{%
Such an analysis will be presented in another paper. The basic idea of this
analysis is to demonstrate that it is highly unlikely that Nature has
invented two drastically different and independent causes of gravitation -
an anisotropic spacetime for elementary charged particles and a curved
spacetime for elementary neutral ones (such as the $Z^{0}$ boson if it turns
out to be a truly neutral particle). If the mass of the elementary charged
particles is regarded as only partly electromagnetic, the phenomenon of
gravitation becomes even more complicated. Spacetime must be anisotropic (to
account for the gravitational interaction of the electromagnetic part of the
mass of the particles) as well as curved (to account for the gravitational
interaction of the non-electromagnetic part of the mass of the charged
particles).}.

\emph{The concept of spacetime curvature is in direct contradiction with
experiments}. A recently obtained result \cite{redshift} shows that the
gravitational redshift contradicts the curved-spacetime interpretation of
general relativity. It has not been noticed up to now that both frequency
and velocity of a photon change in the gravitational redshift experiment. In
such a way the measurement of a change in a photon frequency is in fact an
indirect measurement of a change in its local velocity in this experiment.
This shows that the local velocity of a photon depends upon its pre-history
(whether it has been emitted at the observation point or at a point of
different gravitational potential) - a result that contradicts the standard
curved-spacetime interpretation of general relativity which requires that
the local velocity of light be always $c$ \cite{misner}. Therefore the
gravitational redshift demonstrates that general relativity cannot be
interpreted in terms of spacetime curvature. This situation calls for
another interpretation of the mathematical formalism of general relativity.
Such a possibility of interpreting the Riemann tensor not in terms of
spacetime curvature but in terms of spacetime anisotropy has always existed
since the creation of general relativity but received no attention. In such
an interpretation the Riemannian geometry describes not a curved but an
anisotropic spacetime thus linking gravitation to the anisotropy of
spacetime \cite{phd}.

An additional indication that spacetime around massive bodies is anisotropic
(not curved) comes from the following argument. According to the standard
curved-spacetime interpretation of general relativity the gravitational
effects observed in a non-inertial reference frame $N^{g}$ on the Earth's
surface are caused by the curvature of spacetime originating from the
Earth's mass. The principle of equivalence requires that what is happening
in $N^{g}$ be happening in a non-inertial (accelerating) frame $N^{a}$ as
well. The gravitational effects in general relativity include time and
length effects in addition to the pre-relativistic ones (falling of bodies
and their weight). These, according to the standard interpretation of
general relativity, are also caused by the spacetime curvature around the
Earth. By the principle of equivalence the time and length effects must be
present in $N^{a}$ as well. An inertial observer can explain those effects
happening in $N^{a}$ by employing only special relativity \cite{schiff}. An
observer in $N^{a}$, however, can explain them neither by \emph{directly}
making use of the frame's acceleration nor by the anisotropic velocity of
light there since the concepts of time and space are more fundamental than
the concepts of velocity and acceleration. The non-inertial observer in $%
N^{a}$ has to prove that spacetime in $N^{a}$ is anisotropic due to $N^{a}$%
's acceleration by obtaining the spacetime interval in $N^{a}$. Only then
the observer can derive the time and length effects in $N^{a}$ by using the
anisotropic spacetime interval there. Therefore the standard interpretation
of general relativity leads to a picture involving the principle of
equivalence that is not quite satisfying: if one cannot distinguish between
the effects in $N^{a}$ and in $N^{g}$ then why is the spacetime in $N^{a}$
anisotropic while in $N^{g}$ it is curved. Taking into account the two
results discussed above the picture becomes perfectly consistent: spacetime
in both $N^{a}$ and $N^{g}$ is anisotropic (in $N^{a}$ the spacetime
anisotropy is caused by the frame's acceleration while in $N^{g}$ it
originates from the elementary charges that comprise the Earth).

The result that spacetime is anisotropic (in which the velocity of light is
different in different directions) has enormous implications for both
understanding the nature of inertia and gravitation and the possibility of
controlling them since both inertia and gravitation turn out to be
electromagnetic in origin (at least in part\footnote{%
It is now an established (but unexplainably ignored \cite{pearle}) fact from
the electromagnetic mass theory that the inertial mass and inertia are at
least partly electromagnetic in origin \cite{rohrlich, butler}.}). There
exist two theoretical possibilities for electromagnetic manipulation of
inertia and gravitation:

(i) \emph{Changing the anisotropy of spacetime.} Since one of the
corollaries of the electromagnetic mass theory is that the anisotropy of
spacetime around a body is caused by the body's charged constituents (and
their electromagnetic fields), the employment of strong electromagnetic
fields can create a local spacetime anisotropy which may lead to a body
being propelled without being subjected to a direct force.

(ii) \emph{Using the spacetime anisotropy.} Due to the anisotropic velocity
of light in an accelerating reference frame the electromagnetic attraction
of the opposite charges of an accelerating electric dipole enhances its
accelerated motion - it leads to a self-sustaining accelerated motion
perpendicular to the dipole's axis \cite{cornish}. According to the
principle of equivalence an electric dipole supported in an uniform
gravitational field should levitate \cite{griffiths}. In such a way a strong
electromagnetic attraction between oppositely charged parts of a
non-inertial device may lead to its propulsion or at least to a reduction of
its mass; thus allowing for the external force that accelerates the device
(or the weight of the device) to be reduced. \newline
\indent The possibility of manipulating inertia and gravitation by changing
the anisotropy of spacetime was reported in \cite{petkov}. This paper deals
with the possibility of altering inertia and gravitation by using the
spacetime anisotropy.

\section*{Self-Sustained Acceleration}

The equations of classical electrodynamics applied to an accelerating
electric dipole show that it can undergo self-sustaining accelerated motion
perpendicular to its axis, meaning that not only does the electromagnetic
attraction of the opposite charges of a dipole not resist its accelerated
motion but further increases it. The application of the principle of
equivalence shows that an electric dipole supported in an uniform
gravitational field will be also subjected to a self-sustained acceleration
which may lead to the dipole's levitation. Here we shall derive this effect
in a gravitational field directly without applying the equivalence
principle. \newline
\indent Consider a non-inertial reference frame $N^{g}$ supported in a
gravitational field of strength $\mathbf{g}$. The gravitational field is
directed opposite to the $y$ axis. A dipole with a separation distance $d$
between the two charges is laying along the $x$ axis. Due to the spacetime
anisotropy in $N^{g}$ (manifesting itself in the anisotropic velocity of
light in $N^{g}$) the electric field of the negative charge $-q$ with
coordinates ($d,0$) at a point with coordinates ($0,0$), where the positive
charge $+q$ is, is distorted\footnote{%
This is the electric field of a charge at rest in a gravitational field. If
the charge is uniformly accelerated with $\mathbf{a}=-\mathbf{g}$ its
electric field at a distance $d$ from the negative charge is \cite{jackson,
griffithsb}
\par
\[
\mathbf{E}_{-+}^{a}=\frac{-q}{4\pi \epsilon _{o}}\left( \frac{\mathbf{n}_{-+}%
}{d^{2}}+\frac{\mathbf{a\cdot n}_{-+}}{c^{2}d}\mathbf{n}_{-+}-\frac{\mathbf{a%
}}{c^{2}d}\right) {.}
\]
This is the electric field as described in an inertial reference frame. The
calculation of the electric field in the accelerated frame in which the
dipole is at rest gives the same expression due to the anisotropy of
spacetime in that frame \cite{phd}.} \cite{phd}
\begin{equation}
\mathbf{E}_{-+}^{g}=\frac{-q}{4\pi \epsilon _{o}}\left( \frac{\mathbf{n}_{-+}%
}{d^{2}}-\frac{\mathbf{g\cdot n}_{-+}}{c^{2}d}\mathbf{n}_{-+}+\frac{\mathbf{g%
}}{c^{2}d}\right)  \label{E_g}
\end{equation}
where $\mathbf{n}_{-+}$ is a unit vector pointing from the negative charge
toward the positive charge and $\mathbf{n}_{-+}=-\mathbf{\hat{x}}$ where $%
\mathbf{\hat{x}}$ is a unit vector along the $x$ axis. Since $\mathbf{g\cdot
n}_{-+}=0$ ($\mathbf{g}$ is orthogonal to $\mathbf{n}_{-+}$) the electric
field (\ref{E_g}) reduces to

\[
\mathbf{E}_{-+}^{g}=\frac{q}{4\pi \epsilon _{o}d^{2}}\mathbf{\hat{x}}-\frac{q%
}{4\pi \epsilon _{o}c^{2}d}\mathbf{g}.
\]
The force with which the negative charge attracts the positive one in the
anisotropic spacetime in $N^{g}$ is \cite{phd, petkov}

\[
\mathbf{F}_{-+}^{g}=q\left( 1-\frac{\mathbf{g\cdot n}_{+-}}{2c^{2}}\right)
\mathbf{E}_{-+}^{g}
\]
where $\mathbf{n}_{+-}$ is a unit vector pointing from the positive charge
toward the negative charge. Noting that $\mathbf{g\cdot n}_{+-}=0$ we can
write

\begin{equation}
\mathbf{F}_{-+}^{g}=\frac{q^{2}}{4\pi \epsilon _{o}d^{2}}\mathbf{\hat{x}-}%
\frac{q^{2}}{4\pi \epsilon _{o}c^{2}d}\mathbf{g.}  \label{F-+}
\end{equation}
The first term in (\ref{F-+}) is the ordinary force with which the negative
charge attracts the positive one. The second term represents the vertical
component of the force (\ref{F-+}) that is opposite to $\mathbf{g}$ and has
a levitating effect on the positive charge.

The calculation of the force with which the negative charge of the dipole is
attracted by the positive one gives

\begin{equation}
\mathbf{F}_{+-}^{g}=-\frac{q^{2}}{4\pi \epsilon _{o}d^{2}}\mathbf{\hat{x}-}%
\frac{q^{2}}{4\pi \epsilon _{o}c^{2}d}\mathbf{g.}  \label{F+-}
\end{equation}
The net (self) force acting on the dipole as a whole is directly obtained
from (\ref{F-+}) and (\ref{F+-})

\begin{equation}
\mathbf{F}_{self}^{g}=\mathbf{F}_{-+}^{g}+\mathbf{F}_{+-}^{g}=\mathbf{-}%
\frac{q^{2}}{2\pi \epsilon _{o}c^{2}d}\mathbf{g.}  \label{F_g}
\end{equation}
Therefore, unlike the attraction of the charges of an inertial dipole which
does not produce a net force acting on the dipole, the mutual attraction of
the charges of a dipole in a gravitational field becomes unbalanced and
results in a self-force which opposes the dipole's weight. The effect of the
self-force (\ref{F_g}) on the dipole can be explained in a sense that a
fraction of the dipole of mass
\begin{equation}
m_{att}=\frac{q^{2}}{2\pi \epsilon _{o}c^{2}d},  \label{m_att}
\end{equation}
resulting from the unbalanced attraction of the two charges, is subjected to
an acceleration $-\mathbf{g}$ as long as the dipole stays in a gravitational
field of strength $\mathbf{g}$. While the mass (\ref{m_att}) remains smaller
than the dipole mass the effect of the self-force (\ref{F_g}) will be a
reduction of the dipole mass by $m_{att}$ since the self-force is opposite
to the dipole's weight. When $m_{att}$ becomes equal to the dipole mass
(i.e. when $\mathbf{F}_{self}^{g}$ becomes equal to the weight of the
dipole), the dipole starts to levitate. Further increase of $m_{att}$ will
result in lifting of the dipole.

If the charges of the dipole are an electron and a positron its weight is $%
\mathbf{F}=2m_{e}\mathbf{g}$, where $m_{e}$ is the mass of the electron (and
the positron). Using the electron electromagnetic mass \cite{lorentz}
\begin{equation}
m_{e}=\frac{e^{2}}{4\pi \epsilon _{o}c^{2}r_{0}},  \label{m_e}
\end{equation}
where $r_{0}$ is the classical electron radius, we can calculate the
resultant force acting on the dipole supported in a gravitational field
\begin{equation}
\mathbf{F}_{res}=\mathbf{F+F}_{self}^{g}=\frac{e^{2}}{2\pi \epsilon _{o}c^{2}%
}\left( \frac{1}{r_{0}}\mathbf{-}\frac{1}{d}\right) \mathbf{g.}
\label{F_res}
\end{equation}
As seen from (\ref{F_res}) the dipole will start to levitate when the
separation distance $d$ between its charges becomes equal to $r_{0}$.
However, this could hardly be achieved in a laboratory since $r_{0}\sim
10^{-15}$ m.

Consider now a reference frame $N^{a}$ which is uniformly accelerating with
an acceleration $\mathbf{a}=-\mathbf{g}$. Let the dipole be at rest in $%
N^{a} $ placed in such a way that the acceleration $\mathbf{a}$ is
perpendicular to its axis. In a similar fashion to what we have done in the
case of a dipole in $N^{g}$ here too can be shown that due to the
anisotropic speed of light in $N^{a}$ there is a self-force acting on the
dipole as a whole which is given by

\begin{equation}
\mathbf{F}_{self}^{a}=\frac{q^{2}}{2\pi \epsilon _{o}c^{2}d}\mathbf{a}
\label{F_a}
\end{equation}
(the self-force (\ref{F_a}) can be directly obtained from (\ref{F_g}) by
applying the equivalence principle and substituting $\mathbf{g}=-\mathbf{a}$%
). A fraction of the dipole of mass $m_{att}$ resulting from the unbalanced
attraction of the two charges will be subjected to an acceleration $\mathbf{a%
}$ as long as the whole dipole is experiencing the same acceleration $%
\mathbf{a}$. This means that the fraction of the dipole of mass $m_{att}$
accelerates on its own (due to the unbalanced attraction of the two charges)
which results in a reduction of the dipole's resistance to being accelerated
by the external force. Therefore in order to maintain the same acceleration $%
\mathbf{a}$ the external force accelerating the dipole should be reduced.
Stated another way, the dipole mass is effectively reduced and the
resistance which the dipole offers to being accelerated will reduce as well.
When $m_{att}$ becomes equal to the dipole mass the resistance of the dipole
to being accelerated will cease and consequently there will be no external
force needed to accelerate it. The dipole will continue to maintain its
acceleration entirely on its own - it will be in a state of self-sustained
accelerated motion. This type of motion resembles the inertial motion of an
object: as a free object continues to move with constant velocity until
being prevented from doing so, a dipole (initially accelerated by an
external force) whose charges and separation distance ensure that $m_{att}$
is equal to the dipole mass will continue to move with constant acceleration
on its own until being prevented from doing so.

If the charges of the dipole are an electron and a positron the external
force accelerating the dipole will be
\[
\mathbf{F}_{ext}=2m_{e}\mathbf{a}.
\]
Taking into account (\ref{F_a}) the dipole will mantain a constant
acceleration if
\[
\mathbf{F}_{ext}+\mathbf{F}_{self}^{a}=2m_{e}\mathbf{a.}
\]
Noting that for an electron and a positron the mass in (\ref{F_a}) will be
\[
m_{att}=\frac{e^{2}}{2\pi \epsilon _{o}c^{2}d}
\]
we can write
\[
\mathbf{F}_{ext}=\left( 2m_{e}-m_{att}\right) \mathbf{a}.
\]
If we assume that $m_{att}>2m_{e}$ (which is unlikely to be achieved since
the separation distance $d$ between the charges should be smaller than the
dimension $r_{0}$ of the classical electron considered) an external force
would be needed to slow down the dipole in order that it maintains its
uniform acceleration $\mathbf{a}$.

The effect of mass reduction caused by the mutual attraction of the
accelerating dipole's charges can be described in the following way as well.
Instead of regarding the self-force (\ref{F_a}) as subjecting only a part of
the dipole of mass $m_{att}$ to the acceleration $\mathbf{a}$ it is also
possible to say that $\mathbf{F}_{self}^{a}$ subjects the whole dipole of
mass $2m_{e}$ to an acceleration $\mathbf{a}_{att}$ originating from the
unbalanced attraction between the electron and the positron. Then using the
electron electromagnetic mass (\ref{m_e}) we obtain the relation between $%
\mathbf{a}_{att}$ and $\mathbf{a}$

\[
\frac{e^{2}}{2\pi \epsilon _{o}c^{2}r_{0}}\mathbf{a}_{att}\mathbf{=}\frac{%
e^{2}}{2\pi \epsilon _{o}c^{2}d}\mathbf{a}
\]
or

\begin{equation}
\mathbf{a}_{att}=\frac{r_{0}}{d}\mathbf{a.}  \label{a_att}
\end{equation}
As seen from (\ref{a_att}) the dipole will experience a self-sustained
acceleration $\mathbf{a}_{att}\geq \mathbf{a}$ if $d\leq r_{0}$.

\section*{Conclusion}

At present it does not appear realistic to expect that a self-sustained
acceleration of a body (equal or greater than its initial acceleration) can
be achieved. However, the possibility for eventual practical applications of
the effect of mass reduction can be assessed if macroscopic charge
distributions are considered. It appears that most promising will be the use
of specially designed capacitors like the commercial ones which consist of
alternatively charged layers of metal foil rolled into the shape of a
cylinder with the cylinder axis parallel to $\mathbf{a}$ or $\mathbf{g}$.
Such capacitors can be charged to large amounts of charge. There are
capacitors already available on the market that can carry a charge well
above $1C$. With such (and greater) amounts of charge the experimental
testing of the mass reduction effect appears now possible although it is not
a real issue since this effect is a direct consequence of the classical
electrodynamics when applied to a non-inertial dipole and for this reason
there should be no doubt that it will be experimentally confirmed. The
purpose of this paper is to demonstrate that the practical applicability of
the mass reduction effect may be within reach if proper technological effort
is invested.

\section*{Acknowledgement}

I would like to thank Dr. G. Hathaway for bringing to my attention the
papers of Cornish \cite{cornish} and Griffiths \cite{griffiths} in a
discussion of the effects considered here.

\end{document}